\documentclass[twocolumn,prb,showpacs,floatfix,superscriptaddress]{revtex4-1}

\usepackage{graphicx}
\usepackage{bm}
\usepackage{amsmath}
\usepackage{amsfonts}
\usepackage{amssymb}
\usepackage{xcolor}
\usepackage{float}
\usepackage{xspace}
\usepackage{scalerel}
\usepackage{multirow}
\usepackage{booktabs}
\usepackage{array}

\newcommand{\sg}{$P\,\bar{3}\,c\,1$}

\newcommand{\CTO}{Co$_{4}$Ta$_{2}$O$_{9}$\xspace}
\newcommand{\CNO}{Co$_{4}$Nb$_{2}$O$_{9}$\xspace}
\newcommand{\ABO}{$\mathcal{A}$$_{4}$$\mathcal{B}$$_{2}$O$_{9}$ ($\mathcal{A}$ = Mn, Fe, Co, and $\mathcal{B}$ = Nb, Ta)\xspace}
\newcommand{\ABOshort}{$\mathcal{A}$$_{4}$$\mathcal{B}$$_{2}$O$_{9}$\xspace}

\newcommand{\CO}{Cr$_{2}$O$_{3}$\xspace}

\newcommand{\MTO}{Mn$_{4}$Ta$_{2}$O$_{9}$\xspace}
\newcommand{\FNO}{Fe$_{4}$Nb$_{2}$O$_{9}$\xspace}

\begin{document}
\title{\textbf{Noncollinear antiferromagnetic order in the buckled honeycomb lattice of magnetoelectric \CTO determined by single-crystal neutron diffraction}}

\author{Sungkyun Choi}
\email{sc1853@physics.rutgers.edu}
\affiliation{Department of Physics and Astronomy, Rutgers University, Piscataway, New Jersey 08854, USA}
\affiliation{Max Planck Institute for Solid State Research, Heisenbergstrasse 1, 70569 Stuttgart, Germany}

\author{Dong Gun Oh}
\affiliation{Department of Physics, Yonsei University, Seoul 03722, Korea}

\author{Matthias J. Gutmann}
\affiliation{ISIS Facility, Rutherford Appleton Laboratory, Chilton, Didcot, OX11 0QX, UK}

\author{Shangke Pan}
\affiliation{Department of Physics and Astronomy, Rutgers University, Piscataway, New Jersey 08854, USA}
\affiliation{Rutgers Center for Emergent Materials, Rutgers University, Piscataway, New Jersey 08854, USA}
\affiliation{State Key Laboratory Base of Novel Function Materials and Preparation Science, School of Material Sciences and Chemical Engineering, Ningbo University, Ningbo, Zhejiang 315211, China}

\author{Gideok Kim}
\altaffiliation[Present address: ] {Center for Integrated Nanostructure Physics, Institute for Basic Science (IBS), Suwon 16419, Republic of Korea; Sungkyunkwan University (SKKU), Suwon 16419, Republic of Korea}
\affiliation{Max Planck Institute for Solid State Research, Heisenbergstrasse 1, 70569 Stuttgart, Germany}

\author{Kwanghyo Son}
\affiliation{Max Planck Institute for Intelligent Systems, Heisenbergstrasse 3, 70569 Stuttgart, Germany}

\author{Jaewook Kim}
\altaffiliation[Present address: ] {Korea Atomic Energy Research Institute, Daejeon, Republic of Korea 34057}
\affiliation{Department of Physics and Astronomy, Rutgers University, Piscataway, New Jersey 08854, USA}
\affiliation{Rutgers Center for Emergent Materials, Rutgers University, Piscataway, New Jersey 08854, USA}

\author{Nara Lee}
\affiliation{Department of Physics, Yonsei University, Seoul 03722, Korea}

\author{Sang-Wook Cheong}
\affiliation{Department of Physics and Astronomy, Rutgers University, Piscataway, New Jersey 08854, USA}
\affiliation{Rutgers Center for Emergent Materials, Rutgers University, Piscataway, New Jersey 08854, USA}

\author{Young Jai Choi}
\affiliation{Department of Physics, Yonsei University, Seoul 03722, Korea}

\author{Valery Kiryukhin}
\affiliation{Department of Physics and Astronomy, Rutgers University, Piscataway, New Jersey 08854, USA}

\date{\today}

\begin{abstract}
\CTO exhibits a three-dimensional magnetic lattice based on the buckled honeycomb motif. It shows unusual magnetoelectric effects, including
the sign change and non-linearity. These effects cannot be understood without the detailed knowledge of the magnetic structure.
Herein, we report neutron diffraction and direction-dependent magnetic susceptibility measurements on \CTO single crystals. Below 20.3~K, we find
a long-range antiferromagnetic order in the alternating buckled and flat honeycomb layers of Co$^{2+}$ ions stacked along the $\bm{c}$ axis.
Within experimental accuracy, the magnetic moments lie in the $\bm{ab}$ plane. They form a canted antiferromagnetic structure with a tilt
angle of $\sim$~14$^{\circ}$ at 15~K in the buckled layers, while the magnetic moments in each flat layer are collinear. This is directly evidenced
by a finite (0, 0, 3) magnetic Bragg peak intensity, which would be absent in the collinear magnetic order. The magnetic space group is $C2'/c$.
It is different from the previously reported $C2/c'$ group, also found in the isostructural \CNO. The revised magnetic structure
successfully explains the major features of the magnetoelectric tensor of \CTO within the framework of the spin-flop model.
\end{abstract}
\maketitle

\section{introduction}
\label{sec:intro}
Controlling magnetic order with an electric field, and electric polarization with a magnetic field are of both technological and fundamental significance.
Energy-efficient devices of new types could be developed, for example, using the cross-coupling between the electric and magnetic orders in
magnetoelectric (ME) and multiferroic compounds~\cite{Spaldin2019}. The research of the ME effects was initiated by the theoretical
proposal~\cite{Dzyaloshinskii1960} that
the cross-coupling terms between the electric and the magnetic fields in the
free energy are allowed in the compounds exhibiting certain
structural and magnetic symmetries.
The first compound exhibiting the ME effect, Cr$_2$O$_3$, was discovered shortly
thereafter~\cite{Astrov1960}. Since then, significant efforts were devoted to the search of new ME compounds, especially of those with a strong ME
coupling as relevant to technological applications. This search is hindered by the rather restrictive requirements on the
symmetries for the candidate materials~\cite{Spaldin2019}. However, even when such requirements are met, a physical mechanism producing the strong ME
coupling must be
present in the system. As a result, the number of compounds showing the strong ME effect is still quite limited.

\CO~crystallizes in the corundum structure. It displays a rather strong and linear ME effect, which are both desirable properties. Compounds
possessing similar structures, combined with the
increased spin-lattice coupling, may therefore be considered promising ME candidates. The \ABO~\cite{Bertaut1961} compounds are based on the
corundum structure, and contain magnetic ions such as Co$^{2+}$ and Fe$^{2+}$ expected to have significant orbital magnetic moments conducive
to increased spin-lattice coupling. In addition, they contain heavy nonmagnetic elements (Nb, Ta) with strong on-site spin-orbit interaction,
which may facilitate the ME coupling indirectly. Recently, several members of this family have indeed been shown to exhibit complicated ME and
multiferroic properties ~\cite{Khanh2016, Khanh2017, Panja2018, Maignan2018:FTO, Maignan2018:FNO, Panja2019, Lee2019}.

\begin{figure}[t]
\includegraphics[width=\linewidth]{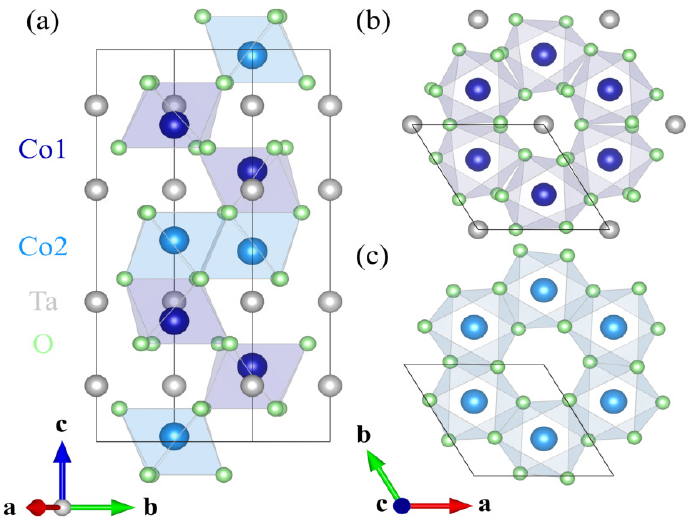}
\caption {(color online) (a) Crystal structure of \CTO. Black solid lines define the unit cell. Crystallographically-different Co1 and Co2 sites
are shown in dark and light blue, respectively. CoO$_6$ octahedra are highlighted. The buckled honeycomb (0.1~$\leq$~$z$~$\leq$~0.4), and the flat
honeycomb (0.4~$\leq$~$z$~$\leq$~0.5) layer fragments are shown in (b) and (c), respectively.
}
\label{fig:str}
\end{figure}

The structure of the \ABOshort~compounds is derived from the well-known corundum crystal structure of \CO~in which four Cr sites are occupied by
the magnetic $\mathcal{A}^{2+}$ ions, and another two sites by the nonmagnetic $\mathcal{B}^{5+}$ ions. This structure is illustrated in
Fig.~\ref{fig:str} using \CTO as an example~\cite{Lee2019}. The space group is trigonal \sg~(No.~165)~\cite{Chaudhary2019}.
There are two crystallographically distinct sites occupied by the Co$^{2+}$ ions, both in the 4$d$ Wyckoff position (1/3, 2/3, $z$):
Co1 at $z\sim$ 0.192, and Co2 at $z\sim$ 0.986~\cite{Lee2019}. Co1 forms the buckled
honeycomb network shown in Fig.~\ref{fig:str}(b) [denoted as the buckled layer in this paper], while Co2 makes a rather flat
honeycomb layer shown in
Fig.~\ref{fig:str}(c), which we call flat for simplicity. These layers are stacked along the $\bm{c}$ axis, forming a three-dimensional structure.

The ME properties of \CNO have been studied most extensively so far. It shows a significant linear ME effect in the magnetically ordered
state~\cite{Khanh2016}, arguably influenced ~\cite{Khanh2016,CNO_soc1,CNO_soc2}
by the unquenched orbital moment of Co$^{2+}$. The direction of the induced electric polarization
can be switched by rotating the magnetic field in the honeycomb plane~\cite{Khanh2017} in the manner consistent with the trigonal symmetry
of the system~\cite{Matsumoto2019}.
Magnetic order is key for understanding the ME effect. However, the magnetic structure of \CNO has been a subject of certain
controversy~\cite{Khanh2016,Deng2018,Ding2020:CNO}. Specifically, it is not agreed whether the magnetic moments are collinear~\cite{Khanh2016} or
tilted~\cite{Deng2018,Ding2020:CNO} within the buckled honeycomb planes, as well as whether any $\bm{c}$ axis moment components are
present~\cite{Khanh2016} or absent~\cite{Deng2018,Ding2020:CNO}. Some of the large components of the ME tensor in \CNO have been accounted for
by the spin-flop model in Ref.\ \onlinecite{Khanh2016}. However, this model does not explain all the tensor components observed experimentally.
An alternative scenario has been proposed in
a more recent neutron diffraction study in an applied magnetic field~\cite{Ding2020:CNO}.
No spin flop was found in these experiments, and the
observed ME effects were proposed to be associated with
the field-dependent magnetic domain populations combined with tilting moments within the domains, at
least in moderate applied fields (below 4 T).
It is clear that detailed knowledge of the magnetic structure, as well as understanding of the magnetic domain
population factors and their effects are needed to explain the complex magnetoelectricity of \CNO.

Recently, intricate ME response has been reported in the magnetically ordered state of \CTO~\cite{Lee2019}. In contrast to the isostructural
\CNO, this response is strongly nonlinear. Also, for some directions of the applied magnetic field, the induced electric polarization changes
its sign with the increasing field. This is somewhat puzzling because the only major difference between \CTO and \CNO is in the non-magnetic ions,
Ta$^{5+}$ and Nb$^{5+}$, respectively. The origin of the distinct ME effects in these two compounds, therefore, deserves a detailed investigation.
The magnetic structure of \CTO has been recently studied by neutron powder diffraction~\cite{Chaudhary2019}.
An $\bm{ab}$-plane noncollinear magnetic
structure similar to the one determined in Refs.\ \onlinecite{Deng2018,Ding2020:CNO} for \CNO,
as well as the same magnetic space group, $C2/c'$, were reported.
Given the current controversial results on the magnetic structure of \CNO, the key role of the magnetic order in the ME effect, and the apparent
failure of the current models to explain this effect in \CNO, it is essential to determine the magnetic structure of \CTO confidently. We also note
that conflicting reports exist for the magnetic structure of the isostructural \FNO~\cite{Jana2019:FNO,Ding2020:FNO},
which further emphasizes the difficulty of the magnetic structure problem in the \ABOshort~compound series.
Single-crystal neutron diffraction studies, supported by other relevant measurements, are essential to address this problem.

In this paper, we determine the magnetic structure of \CTO using single-crystal neutron diffraction complemented by direction-dependent magnetic
susceptibility measurements. Some characteristic features of the magnetic structure of \CTO found in our study, such as
the $\bm{ab}$-plane magnetic moments and
the canted (collinear) antiferromagnetic order in the buckled (flat) honeycomb layers, are similar to those reported in the
previous work on \CTO~\cite{Chaudhary2019}
and the isostructural \CNO~\cite{Deng2018,Ding2020:CNO}. However, we find a distinct magnetic space group, $C2'/c$, as opposed to $C2/c'$ reported
in these references. The major difference is the direction of the refined magnetic moments. Symmetry analysis of the newly-determined magnetic order of
\CTO accounts for the observed ME tensor~\cite{Lee2019} in the framework of the spin-flop model~\cite{Khanh2016}, giving a better agreement with
the experimental data than the previously-reported magnetic space group. While the neutron diffraction results do favor the $C2'/c$ over the
$C2/c'$ magnetic space group, the difference in the reliability factor is small. The direction-dependent magnetic susceptibility data were crucial
for the confirmation of the $C2'/c$ magnetic space group in \CTO. Our results emphasize the importance of the careful choice of the
complementary measurements for the determination of the correct magnetic structure of the \ABOshort~compounds.

This paper is organized as follows. Section~\ref{sec:exp} describes the experimental details. Magnetic susceptibility and neutron diffraction results are
given in Section~\ref{sec:dc:ac} and Section~\ref{sec:ND}, respectively. Implications of these results are discussed in Section~\ref{sec:discussion},
and conclusions are given in Section~\ref{sec:summary}. Appendix~\ref{app:sec:az} and Appendix~\ref{app:ND} provide details of the analysis of
the magnetic susceptibility and neutron diffraction data, respectively. The symmetries of the $C2'/c$ and the $C2/c'$ magnetic space groups and the
corresponding magnetic structures are discussed in Appendix~\ref{app:MSG}, and the magnetic structure factors of the (0, 0, L) magnetic Bragg peaks
are given in Appendix~\ref{app:MSF:001}.

\begin{figure}[t]
\includegraphics[width=\linewidth]{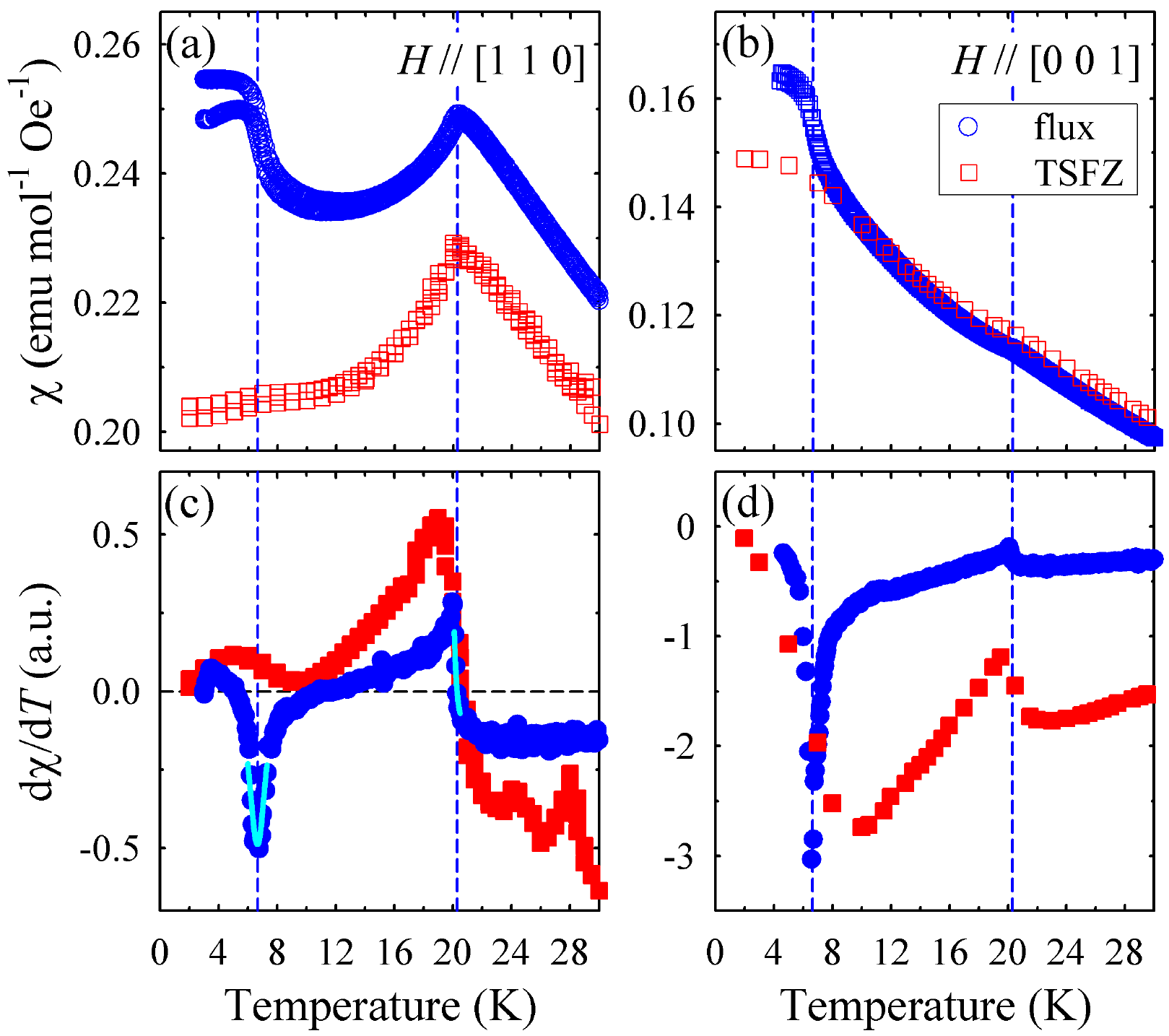}
\caption {(color online) ZFC and FC magnetic susceptibilities for the magnetic field along (a) the [1 1 0] and (b) [0 0 1] trigonal directions for the flux (blue empty
circles) and TSFZ (red empty squares) crystals. The FC data for the TSFZ crystal are omitted in (b). The corresponding temperature
derivatives for the ZFC data are shown in (c) and (d). Cyan solid lines indicate the fits used to determine the magnetic transition temperature
at 20.3~K and the magnetic anomaly at 6.65~K. These temperatures are depicted by the vertical dashed lines.
}
\label{fig:dc:ac}
\end{figure}

\begin{figure}
\includegraphics[width=\linewidth]{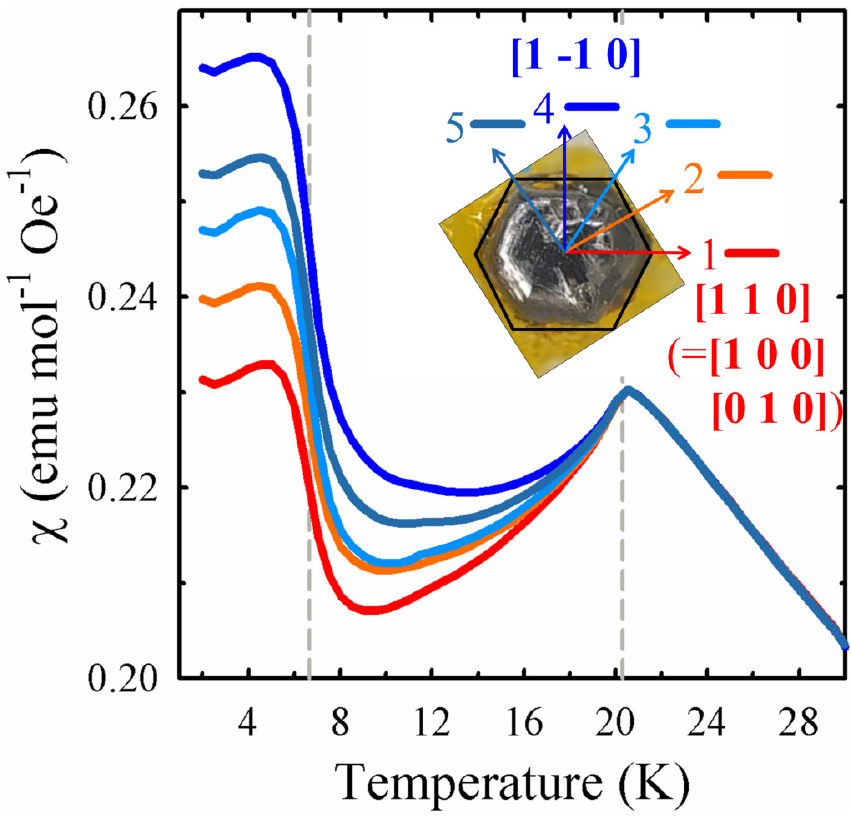}
\caption {(color online) ZFC magnetic susceptibility for various directions of the magnetic fields in the $\bm{ab}$ plane. The data were taken
using the flux-grown crystal depicted in the inset (a different sample from the one characterized in Fig.~\ref{fig:dc:ac}). The inset shows the
color code
for the directions of the magnetic field with respect to the trigonal crystallographic axes, and lists several directions equivalent by the symmetry.
The data were averaged as discussed in the text. Vertical dashed lines indicate the two characteristic temperatures
extracted from the data of Fig.~\ref{fig:dc:ac}(c).
}
\label{fig:SM:rot:Chi}
\end{figure}

\section{Experimental Details}
\label{sec:exp}
Two sets of \CTO single crystals were used in this work. One set was grown using the flux method, as described in Ref.\ \onlinecite{Lee2019}.
We refer to them as the flux crystals. The other set was produced using the traveling solvent floating zone method, the TSFZ crystals.
The polycrystalline powder of \CTO was obtained by a solid-state reaction technique from the stoichiometric mixture of Co$_{3}$O$_{4}$ (99.99\%)
and Ta$_{2}$O$_{5}$ (99.99\%) powders sintered at 1200~$^{\circ}$C for 10 hours in air. It was used to make the feed and seed rods for the TSFZ
growth.

Magnetic susceptibility measurements were done with SQUID magnetometry, using either a normal DC accessory or a reciprocating sample measurement
system to increase the signal to noise ratio. Susceptibility measurements between 30~K and the base temperature (typically 2 or 3 K) were done
in 0.1~T applied magnetic field. Measurements at higher temperatures were also taken. Zero-field-cooled (ZFC) and field-cooled (FC) measurements
were done when necessary. Crystallographic axes were pre-determined by Laue x-ray diffraction, and cross-checked by a fixed-wavelength single-crystal
x-ray diffraction (a Mo source).

Neutron diffraction experiments were performed at the Single Crystal Diffractometer (SXD) beamline at ISIS, where the time-of-flight Laue technique
is used to access large 3-D volumes of reciprocal space in a single measurement. Single crystals were screened by magnetic susceptibility, x-ray
diffraction, followed by further quality checks with neutron Laue diffraction at room-temperature on SXD. One crystal from each growth method
was selected for the neutron diffraction measurements. The 22-mg flux crystal was of a spherical shape, about 2~mm in diameter. The data were collected
at three identical rotation angles, in the paramagnetic state at 25~K for 18 hrs, and in the magnetically ordered state at 15 K for 42 hrs. The
1.28-g TSFZ crystal was of a cylindrical shape, 6~mm in diameter, cut from a bigger rod. For this crystal, the data were collected at six identical
rotation angles at 25 and 15~K for 41.2 and 68.7~hrs, respectively.

\begin{figure*}
\includegraphics[width=\linewidth]{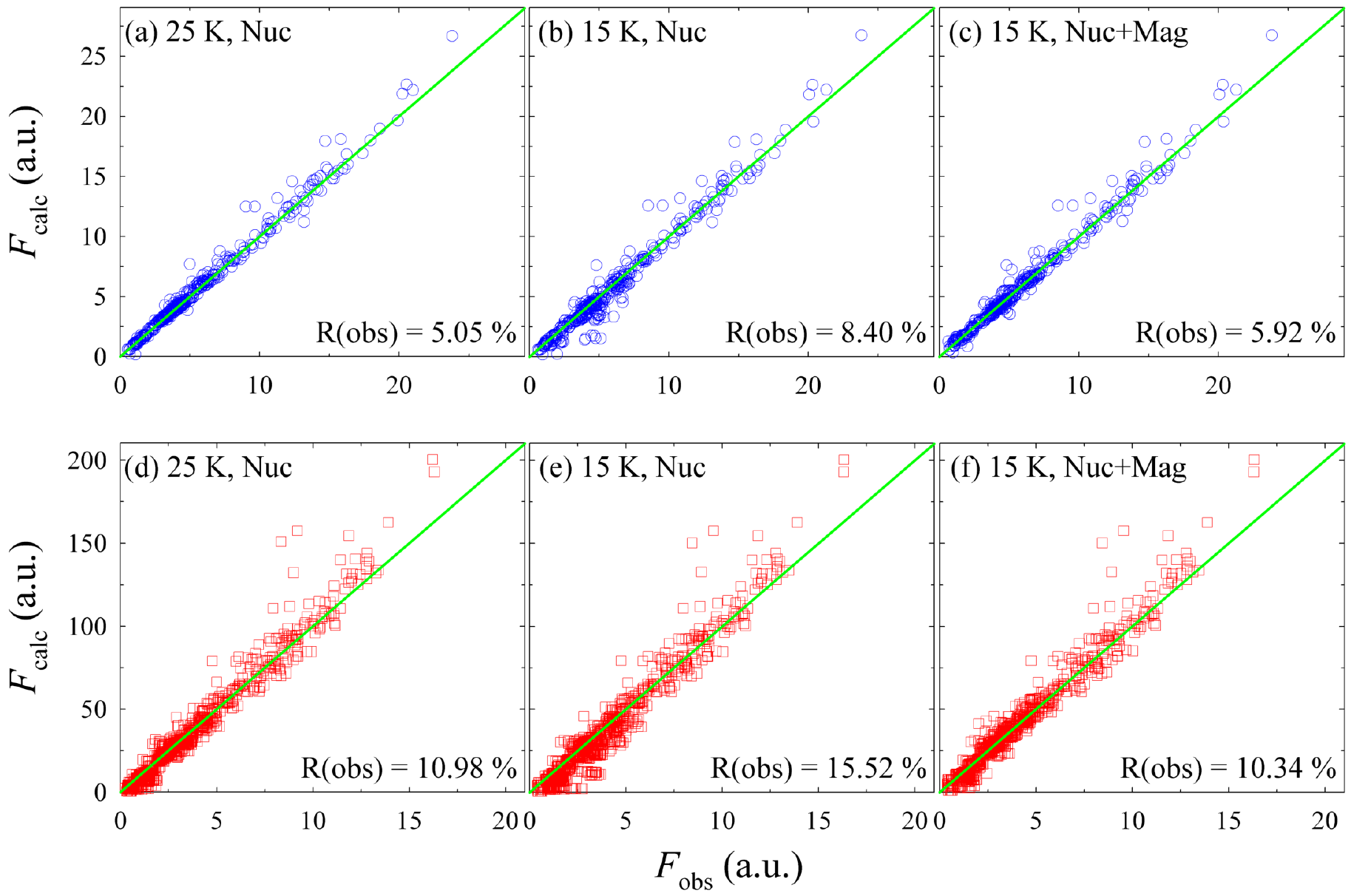}
\caption {(color online) Neutron diffraction data analysis showing the calculated structure factor \textit{F}$_{\rm{calc}}$ versus the observed structure
factor \textit{F}$_{\rm{obs}}$. Panels (a) and (d) are for the paramagnetic state at 25~K, the remaining panels show the data for the ordered state at 15~K.
The observed Bragg peaks with wave vectors $Q\leq$~6.28 \AA$^{-1}$ are shown.
The nuclear structure (Nuc) refinement obtained
using the whole-$Q$ data set at 25 K is compared to the experimental data at 25 K in panels (a) and (d), and at 15 K in panels (b) and (e).
Magnetic refinement results with the fixed nuclear structure (Nuc+Mag) are shown in (c) and (f) for the $C2'/c$
in-plane noncollinear magnetic models marked with the dagger ($\dagger$) symbols in Table~\ref{tab:mag:ref} (see the text for the details).
The top (a-c) and the bottom (d-f) panels are for the flux and the TSFZ crystal, respectively.
}
\label{fig:SM:ND:flux}
\end{figure*}

\section{Magnetic susceptibility}
\label{sec:dc:ac}
We performed direction-dependent magnetic susceptibility measurements on \CTO using both the flux and TSFZ crystals. The major features are shown in
Fig.~\ref{fig:dc:ac}. The in-plane susceptibility with the magnetic field along the [1 1 0] direction is depicted in Fig.~\ref{fig:dc:ac}(a), and
the $\bm{c}$-axis susceptibility (field parallel to [0 0 1]) is shown in Fig.~\ref{fig:dc:ac}(b). Figs.~\ref{fig:dc:ac}(c, d) show the corresponding
derivatives with respect to temperature for the ZFC data. The transition temperature for the antiferromagnetic order \textit{T}$_{\rm N}$, defined from
the sharp cusp of $\chi (T)$ in the in-plane data, is 20.3~K in the both samples. The anomaly at the \textit{T}$_{\rm N}$ is much less
pronounced in the $\bm{c}$-axis data. This means that magnetic moments are dominantly confined in the $\bm{ab}$ plane.

There is an additional anomaly at a lower temperature, at which a bifurcation between the FC and ZFC data is observed. As determined by the minimum
in the temperature derivative, it occurs at 6.65~K in the flux crystal. While less pronounced, this anomaly is also present in the TSFZ crystal at a
slightly higher temperature.
The observed temperature hysteresis and the sample dependence indicate a complex nature of the magnetic state at the lowest temperatures, where
magnetic domain or glassy effects could play a role.
In this work, the \CTO magnetic structure was determined at 15~K, safely outside of this complex regime.
Detailed studies of the magnetism below 6.65~K would be desirable, but are beyond the scope of this paper.

To characterize the in-plane magnetic anisotropy, we made magnetic susceptibility measurements for a large number of representative high-symmetry
directions in the trigonal $\bm{ab}$ plane. A flux crystal of a hexagonal shape with well-defined facets was used. The crystallographic axes
were confirmed by x-ray Laue measurements. Fig.~\ref{fig:SM:rot:Chi} shows the data for five such directions, averaged for several repetitive
measurements as discussed in detail in Appendix~\ref{app:sec:az}. Below \textit{T}$_{\rm N}$, the largest and the lowest susceptibilities are found for the
magnetic field along the [1 -1 0] and the [1 1 0] trigonal directions, respectively. For the other directions, the data interpolate between these
values. These measurements clearly demonstrate that the magnetic moments point predominantly along the [1 1 0] direction, assuming a (nearly)
collinear antiferromagnetic structure and a dominant single magnetic domain state (to be discussed in detail later). By convention, this means that
the magnetic easy axis is [1 1 0], while the hard axis is [1 -1 0]. These results will play an important role in the analysis of the neutron
diffraction data discussed in the next Section.

\begin{table*}
\caption{Magnetic structure refinement results for various models, for the flux and TSFZ samples. The data were collected at T = 15 K.
The observed peaks with intensities $I > 3.0 \times \sigma(I)$ and $Q\leq$~6.28 \AA$^{-1}$~were used in the fits. The single-domain model was used for the
flux sample, while the model with three equally populated magnetic domains was utilized for the TSFZ crystal. Asterisk ({*}) symbols mark the
refinements with unrealistically large moments along the $\bm{c}$ axis, as compared to the moments in the $\bm{ab}$ plane. Dagger ($\dagger$) symbols
indicate the final models. $\mid$${\bm M}$$\mid$ is the magnitude of the magnetic moment. Collinear and noncollinear refer to the arrangements of
the magnetic moments in the $\bm{ab}$ plane.
}
\label{tab:mag:ref}
\begin{center}
\begin{tabular}
[c]{c|c|cc|ccc}
\hline\hline
                                       & \multirow{2}{*}{~~~~~~Magnetic model~~~~~~}        &\multicolumn{2}{|c|}{~~~~~~$C2/c'$ (No.~15.88)~~~~~~}                    & \multicolumn{2}{|c}{~~~~~~$C2'/c$ (No.~15.87)~~~~~~}\\
                                       &                                                                            & R$_{\rm obs}$           & $\mid$${\bm M}$$\mid$ (Co1) / (Co2)                       & R$_{\rm obs}$           & $\mid$${\bm M}$$\mid$ (Co1) / (Co2)  \\
\hline
\multirow{4}{*}{Flux}       & Collinear, M$_{c}$=0                  &~5.94~\%          &~2.40~/~1.34~$\mu_{B}$             &~6.15~\%          &~1.64~/~1.64~$\mu_{B}$ \\
                                       & Collinear, M$_{c}$$\neq$0        &~5.97~\%          &~2.34~/~1.36~$\mu_{B}$$^{*}$    &~6.15~\%          &~1.65~/~1.69~$\mu_{B}$ \\
                                       & Noncollinear, M$_{c}$=0            &~6.31~\%          &~1.89~/~1.89~$\mu_{B}$             &~5.92~\%          &~2.02~/~1.44~$\mu_{B}$$^{\dagger}$ \\
                                       & Noncollinear, M$_{c}$$\neq$0  &~6.06~\%          &~2.11~/~1.63~$\mu_{B}$$^{*}$    &~5.91~\%          &~2.02~/~1.45~$\mu_{B}$ \\
\hline
\multirow{4}{*}{TSFZ}  & Collinear, M$_{c}$=0                   &~10.43~\%        &~2.82~/~1.84~$\mu_{B}$              &~10.71~\%        &~2.20~/~2.20~$\mu_{B}$ \\
                                       & Collinear, M$_{c}$$\neq$0        &~10.31~\%        &~2.82~/~1.84~$\mu_{B}$             &~10.26~\%         &~2.20~/~2.78~$\mu_{B}$$^{*}$ \\
                                       & Noncollinear, M$_{c}$=0            &~10.53~\%        &~2.35~/~2.35~$\mu_{B}$             &~10.34~\%         &~2.74~/~2.04~$\mu_{B}$$^{\dagger}$ \\
                                       & Noncollinear, M$_{c}$$\neq$0  &~10.27~\%        &~2.52~/~2.17~$\mu_{B}$$^{*}$    &~10.03~\%         &~2.64~/~2.60~$\mu_{B}$$^{*}$ \\
\hline\hline
\end{tabular}
\end{center}
\end{table*}

\section{Neutron diffraction}
\label{sec:ND}
To determine the magnetic structure of \CTO, neutron diffraction measurements were performed using the both types of single crystals.
The same flux sample, and TSFZ sample from the same growth were used for the neutron diffraction and the magnetic susceptibility measurements shown in Fig. \ref{fig:dc:ac}.
In this paper, we determine the magnetic structure at T~=~15~K, well above the magnetic transition to
the complex phase at 6.65~K. The results of the neutron diffraction data analysis are shown in Fig.~\ref{fig:SM:ND:flux}. Panels (a-c) and
(d-f) present the calculated versus the observed structure factors for the flux and the TSFZ crystal, respectively.

The nuclear structure in the paramagnetic state was refined first, using the whole data set collected at 25 K. Figs.~\ref{fig:SM:ND:flux}(a, d)
show the results of this refinement for the set of the lower-$Q$ Bragg peaks, as relevant to the magnetic refinements described below.
The crystal structure nearly identical to the one determined at room temperature by x-rays~\cite{Lee2019} was found, in the same \sg~trigonal
space group. When this fixed nuclear structure is used for the data taken in the magnetically ordered state at 15 K, additional unaccounted
diffraction signal is clearly visible for a set of the experimental Bragg peaks. That is, \textit{F}$_{\rm{obs}}$ is larger than \textit{F}$_{\rm{calc}}$ for these peaks.
They are easily seen below the \textit{F}$_{\rm{obs}}$~=~\textit{F}$_{\rm{calc}}$ line in Figs.~\ref{fig:SM:ND:flux}(b, e).
Importantly, this additional intensity is only prominent for the reflections with lower wave vectors $Q$.
This indicates the magnetic origin of the extra intensity for these peaks. In Fig.~\ref{fig:SM:ND:flux},
the same set of Bragg peaks is shown in every panel. A thorough search for any additional Bragg peaks covering both the commensurate and
incommensurate positions in the reciprocal space was carried out in the magnetically ordered phase using our area detector data.
Only a very small number of new Bragg peaks was found (to be discussed later).
They were all indexed using the integer (H, K, L) positions in the parent crystal structure. This shows that the magnetic and the
nuclear structure have the same unit cell, in agreement with the previously published neutron diffraction work~\cite{Chaudhary2019}.
Consequently, the magnetic ordering wave vector is $q$~=~(0, 0, 0).  The set of the observed peaks
shown in Fig.~\ref{fig:SM:ND:flux}, with
$Q\leq$~6.28~\AA$^{-1}$ ~as appropriate for the magnetic structure determination, was used in the magnetic refinements. The nuclear structure was fixed.
The most reliable results,
obtained as discussed below, are shown in Figs.~\ref{fig:SM:ND:flux}(c, f).
The addition
of the magnetic structure to the refinements resulted in the obviously better fits. This is reflected by the reduction of the reliability factor
R$_{\rm obs}$ by 2.48~\% and 5.18~\% for the flux and the TSFZ sample, respectively. The full details of the refinement procedure can be found in
Appendix~\ref{app:ND}.

To find the magnetic space group candidates, group symmetry analysis~\cite{Aroyo2011} was applied to the parent nuclear space group \sg~(No.~165)
with $q$~=~(0, 0, 0). All the trigonal subgroups were inconsistent with the magnetic susceptibility data because they disallowed magnetic moments in
the $\bm{ab}$ plane. The next available highest-symmetry choice is monoclinic. We found two candidates that were compatible with the predominantly
in-plane magnetic moments, and that resulted in good fits to the neutron diffraction data collected at 15~K. These were $C2/c'$ (No.~15.88) and
$C2'/c$ (No.~15.87). One of them, $C2/c'$, was previously proposed for both \CNO~\cite{Khanh2016,Deng2018,Ding2020:CNO} and
\CTO~\cite{Chaudhary2019}. Both these groups allow collinear and noncollinear antiferromagnetic order in the $\bm{ab}$ plane, as well as an
antiferromagnetic $\bm{c}$-axis component. We have carried out refinements of the magnetic structure, starting with the simplest possible model,
the collinear in-plane antiferromagnetic structure with no $\bm{c}$ axis components. More complex models, allowing noncollinear in-plane structures,
as well as the out-of-plane moment, were then considered. When the symmetry is lowered from the trigonal to monoclinic at the magnetic transition,
three types of the in-plane magnetic domains are possible, distinguished by the three possible directions of the unique $\bm{b_m}$ axis in the monoclinic
cell.
We found that the refinements using a single-domain model for the flux crystal worked well, while the model using three equally populated domains
was necessary for the TSFZ crystal.
The multi-domain model was also tested for the flux crystal data. It showed a similarly good
fit with only a marginal improvement. Thus, we conclude that the larger TSFZ crystal contains all the possible magnetic domains
with similar populations, while the
smaller flux crystal probably exhibits a predominant single-domain state.

\begin{figure}
\includegraphics[width=\linewidth]{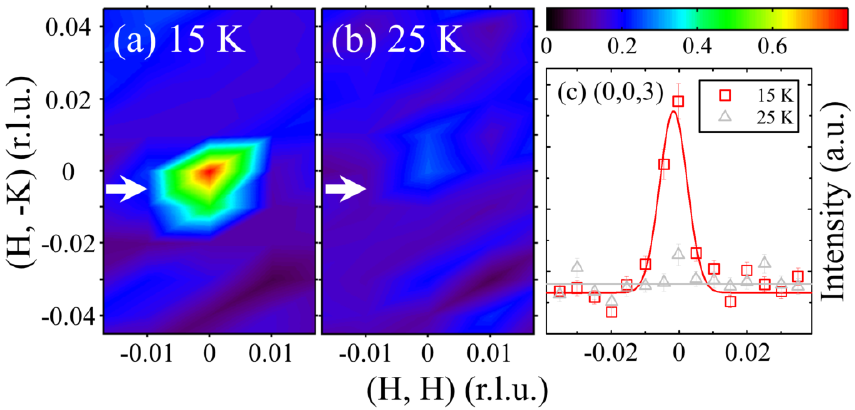}
\caption {(color online) Neutron diffraction patterns in the vicinity of the (0, 0, 3) position in the antiferromagnetic phase at (a) T~=~15~K,
and in the paramagnetic state at (b) T~=~25~K. Panel (c) shows cuts through the peak position in the directions indicated by the white arrows in (a, b)
[2.8 $\leq$ L $\leq$ 3.2, -0.2 $\leq$ (H, -K) $\leq$ 0.1]. The
indexing is done using the trigonal unit cell. A clear magnetic
Bragg peak is observed at 15 K. The slight signal at 25 K is
probably due to multiple scattering. The data are for the TSFZ
crystal.
}
\label{fig:SM:ND:003}
\end{figure}

The refinement results are summarized in Table~\ref{tab:mag:ref}. The fits are characterized by the standard reliability factor R$_{\rm obs}$.
Lower values of R$_{\rm obs}$ signify a better model fit to the data. The difference in the average R$_{\rm obs}$ for the flux and the TSFZ crystal may result from
both the better crystallographic quality, and the smaller number of the observed Bragg peaks for the flux crystal. The main observation from the analysis
given in Table~\ref{tab:mag:ref} is the very small variation of the R$_{\rm obs}$ values for all the models considered for a given crystal. Thus, it appears
impossible to determine reliably the actual magnetic structure based on the fit quality alone. This could be the major reason for the difficulty of
determining the magnetic structure of the \ABOshort~compounds described in the Introduction. Thus, specific signatures of the magnetic order type
should be sought in the diffraction data, and in their absence, complementary experimental techniques should be exploited.

Firstly, in Table~\ref{tab:mag:ref}, we can exclude all the models with the collinear in-plane structures because new Bragg peaks of the (0, 0, odd) type
appear below \textit{T}$_{\rm N}$ at small scattering wave vectors (signifying the magnetic origin of these chemically-forbidden peaks). As an example, the
temperature dependence of the (0, 0, 3) peak is shown in Fig.~\ref{fig:SM:ND:003}. The intensity of these peaks is zero both in the
parent space group, and for any type of the collinear in-plane magnetic order in $C2/c'$ and $C2'/c$, with or without the $\bm{c}$ axis
magnetic moment.
They acquire intensity as the moments start tilting in the $\bm{ab}$ plane, away from the collinear antiferromagnetic alignment.
A detailed discussion of the magnetic structure factor of the (0, 0, L) peaks can be found in Appendix ~\ref{app:MSF:001}.
Secondly, most of the models involving nonzero out-of-plane moment (M$_c$) produced unrealistically large M$_{c}$ values,
which is incompatible with the
magnetic susceptibility results. These models are marked with asterisk symbols in Table~\ref{tab:mag:ref}. The only noncollinear in-plane structure that
converged well with reasonable components M$_{c}$ is the $C2'/c$ model for the flux crystal.
It gives M$_{c}$~$\sim$~0.2~$\mu_B$,
corresponding to the 6$^\circ$ out-of-plane tilt, only exhibited by the flat-layer moments. However, it does not result in any meaningful
improvement of the R$_{\rm obs}$ value over the fully in-plane magnetic structure. We therefore conclude that within the error of our experimental method,
the magnetic moments of \CTO are confined to the $\bm{ab}$ plane. This model is adopted in our paper.
We note that this conclusion matches the results reported in
Ref.\ \onlinecite{Chaudhary2019}~for \CTO, and in Refs.\ \onlinecite{Deng2018,Ding2020:CNO} for the isostructural \CNO.

Among the noncollinear structures with the $\bm{ab}$-plane magnetic moments, our fits slightly favor the $C2'/c$ space group that now has a
meaningfully smaller R$_{\rm obs}$~=~5.92 (10.34)~\% than the R$_{\rm obs}$~=~6.31 (10.53)~\% for the $C2/c'$ space group in the flux (TSFZ) crystal.
However, given the reports of the $C2/c'$ structure for both \CTO~\cite{Chaudhary2019} and \CNO~\cite{Khanh2016,Deng2018,Ding2020:CNO}, a stronger
evidence is needed to determine the magnetic structure confidently.
Such evidence comes from the direction-dependent magnetic susceptibility data. To interpret these data, one must understand the key difference
between two magnetic space groups.
To make comparison to the results of the ME and the magnetic susceptibility
measurements, we use the trigonal notation here and below
in this paper. See Appendix C and
Table II for the full description of the relevant symmetry
operators, as well as for the conversion between the trigonal and the monoclinic axes.
In the collinear in-plane order, for both the Co1 and Co2 sites, magnetic moment (M) components are constrained
by M$_{a}$ = 2 M$_{b}$ for $C2/c'$, while M$_{a}$ = 0 in $C2'/c$.
This means that in the $C2'/c$ group, the spins point along
the [0 1 0] trigonal direction. By symmetry, [1 0 0] and [-1 -1 0] are equivalent to [0 1 0]. Therefore, three types of the in-plane magnetic
domains, with the spins pointing along either [1 0 0], [0 1 0], or [1 1 0] may form. In the $C2/c'$ group, the spin directions are perpendicular
to the ones listed above. One such direction (out of three) is [1 -1 0]. These conclusions may only be slightly modified for the noncollinear states
with a small tilting of the magnetic moments, as is the case here (see below).
Thus, the $C2/c'$ group implies the [1 -1 0] or equivalent easy axis (the major spin
direction) in the magnetic susceptibility measurements, while the easy axis should be along the [1 1 0] or equivalent direction for the $C2'/c$
symmetry. Importantly, the susceptibility measurements should be done in a sample with a nearly-single magnetic domain, otherwise the in-plane
susceptibility averages out. Our direction-dependent susceptibility measurements clearly show that the \CTO flux samples tend to
produce the state with a dominant magnetic domain at low temperatures. This is fortuitous, considering that the TSFZ sample exhibits a multi-domain state.
The direction-dependent susceptibility data shown in Fig.~\ref{fig:SM:rot:Chi} clearly demonstrate that the easy axis is [1 1 0], and the hard axis is [1 -1 0].
The combined neutron diffraction and magnetic susceptibility data therefore unambiguously identify $C2'/c$ as the space group of the magnetic order
in \CTO at 15~K.

\begin{figure} [t]
\includegraphics[width=\linewidth]{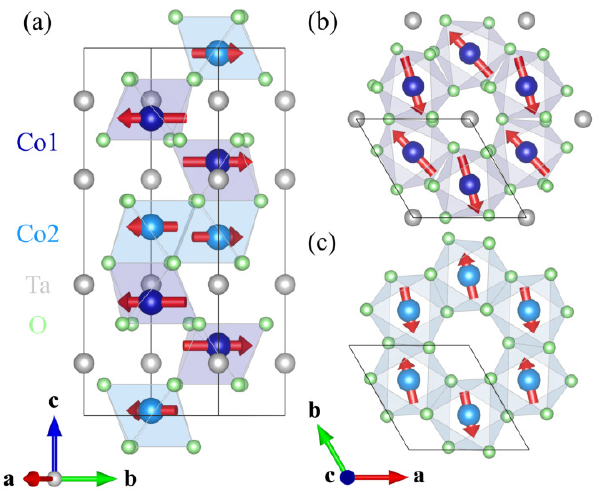}
\caption {(color online) (a) The refined magnetic structure of \CTO at T~=~15~K. The magnetic space group is $C2'/c$. (b) and (c) show the
buckled and the flat honeycomb layer, respectively. The magnetic arrangement is noncollinear (canted antiferromagnetic) in the former, and collinear
antiferromagnetic in the latter. The net moments of the two buckled layers in the unit cell add to zero.
The trigonal crystallographic axes, as well as the corresponding unit cells, are shown. This structure corresponds to the
magnetic domain with the [0 1 0] easy axis.
}
\label{fig:SM:ND:str}
\end{figure}

The refined $C2'/c$-type noncollinear magnetic order in \CTO is shown in Fig.~\ref{fig:SM:ND:str}. The TSFZ crystal refinement,
marked with the dagger ($\dagger$) in Table~\ref{tab:mag:ref} is chosen for this figure. The magnetic structure obtained in the flux crystal
refinement is visually identical. The magnetic domain with the [0 1 0] easy axis
is shown, following the standard crystallographic convention for the conversion from the trigonal to the monoclinic unit cell. The moments are confined
in the $\bm{ab}$ plane. The refined magnetic moment vectors for the Co1 and Co2 sites are, respectively, (-0.73, 2.3, 0)$\mu_B$ and
(0.73, 2.3, 0)$\mu_B$ for the TSFZ crystal, and (-0.61, 1.64, 0)$\mu_B$ and (0.61, 1.64, 0)$\mu_B$ for the flux crystal.
Note, since the trigonal axes utilized in this paper are nonorthogonal, the magnitudes of the refined Co1 and Co2 moments are different,
as listed in Table I.
The other magnetic moments are generated according to the symmetry rules listed in Table~\ref{tab_sym}.
See Appendix B for the detailed discussion of the conditions utilized in the refinement of the moment vectors.
The buckled honeycomb layers exhibit a noncollinear canted antiferromagnetic order, as depicted in Fig.~\ref{fig:SM:ND:str}(b).
The refined tilt angle (from the collinear condition) is 13.35$^\circ$ and 15.20$^\circ$ for the TSFZ and the flux crystal, respectively.
This angle is significantly larger than the 6-7$^\circ$ angle obtained in the first-principles studies of Ref. \onlinecite{Solovyev2016}, probably
due to the overestimation of isotropic exchange interactions.
Because of this tilt, the buckled layer
has a net magnetic moment within one plane. This moment is canceled by the opposite moment of the second buckled layer in the unit cell.
Each flat layer displays the collinear antiferromagnetic order, as shown in Fig.~\ref{fig:SM:ND:str}(c).
The magnetic moments of the nearest neighbors along the $\bf{c}$ axis, one from the buckled and the other from the flat layer, are nearly in the same direction.
As a result, the easy axes of the two flat layers in the unit cell are slightly misaligned, following the tilted moments in the buckled layers.
Because of the small value of the tilt angle in the honeycomb layers, the easy and the hard magnetic axes essentially retain the directions
characteristic to the collinear order. All these features of the magnetic order are determined by the $C2'/c$ symmetry
(see Table~\ref{tab_sym} and Fig.~\ref{fig:SM:sym} in the Appendix for further details).

\section{Discussion}
\label{sec:discussion}
In this paper, we report the refined magnetic order in \CTO. The magnetic space group is $C2'/c$, the magnetic easy axis points
along the trigonal [1 1 0] (or equivalent) direction.
This finding could only be reached by combined single-crystal neutron diffraction and direction-dependent magnetic susceptibility measurements.
A different magnetic space group, $C2/c'$, with the easy axis along [1 -1 0] was reported for the isostructural \CNO
compound~\cite{Khanh2016, Deng2018,Ding2020:CNO}. This difference is intriguing because it highlights a possible significance of the nonmagnetic
ions (Ta$^{5+}$ or Nb$^{5+}$) in the anisotropy of the magnetic Co$^{2+}$ lattice. Both Ta and Nb are heavy
elements with sizeable on-site spin-orbit coupling. Co$^{2+}$ is among the small number of the 3$d$ ions showing a significant orbital component of the
magnetic moment. In combination, these properties may result in a complex coupling between the lattice and magnetic moments. They could therefore
play a key role in the observed complex magnetoelectric effects~\cite{Khanh2016,Khanh2017,Lee2019}.
Systematic first-principles studies are highly desired to establish the origin of
the magnetic anisotropy and the nature of the ME coupling in these compounds. Experimentally, it is important to determine the microscopic
magnetic interactions, as they stabilize the long-range magnetic order, and may help understand the observed ME effect. This is best done using
inelastic neutron scattering. \CNO has been studied using this technique~\cite{Deng2018}, but higher energy resolution is needed to determine the
interactions responsible for the noncollinear structure and for the magnetocrystalline anisotropy.
Such interactions typically depend on the spin-orbit coupling, and therefore are
expected to have an important influence on the ME effect. A well-known example is the Dzyaloshinskii-Moriya interaction~\cite{DM_ref}, discussed in
connection to the \CNO magnetic properties~\cite{Deng2018}.

\begin{figure}
\includegraphics[width=\linewidth]{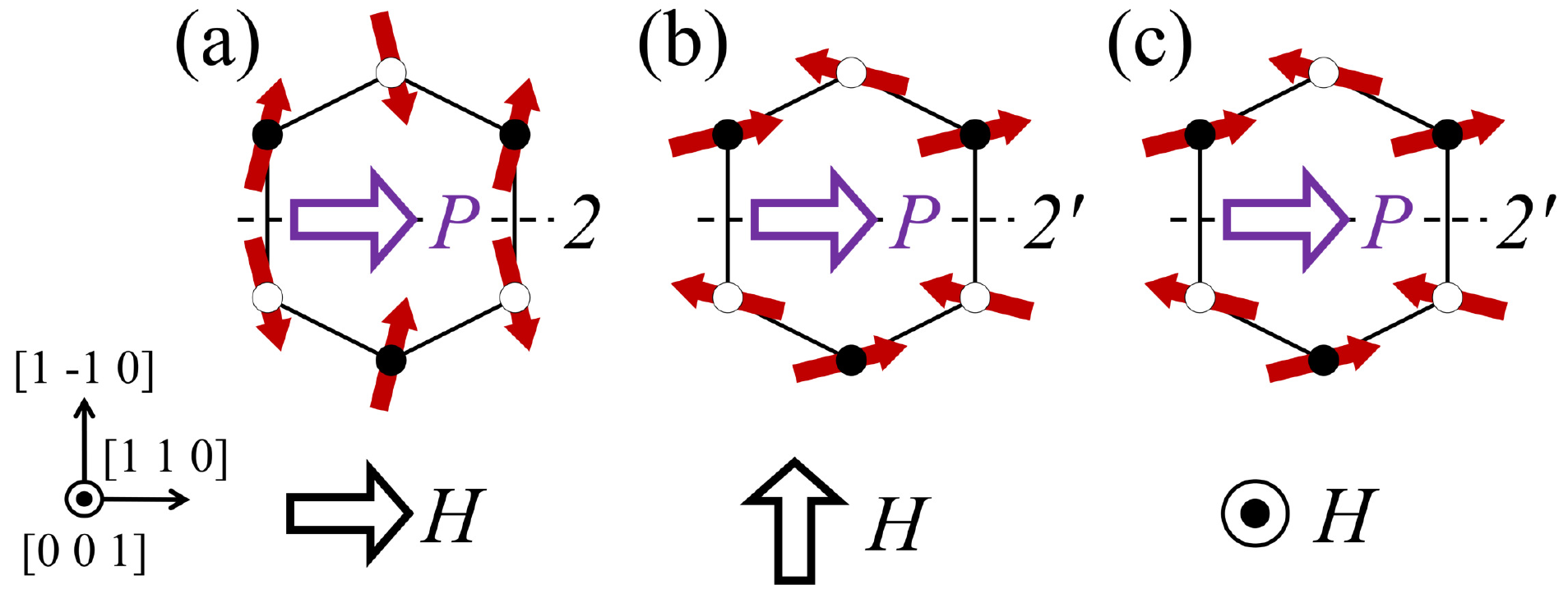}
\caption {(color online) The electric polarization $P$ induced in an applied magnetic field $H$ in \CTO, as predicted by the model described in
the text. Dark red arrows represent the magnetic moments in the buckled honeycomb layer.
Open and closed circles represent the Co sites with two different crystallographic $z$ coordinates in a single layer.
The domain with the [1 1 0] magnetic easy axis is shown in (c).
The magnetic moments in (c) have, in addition, a small out-of-plane component (along the magnetic field direction).
Crystallographic directions are trigonal. Note, no electric polarization is allowed by symmetry in the plane perpendicular to the two-fold
(2 or 2$'$) axes shown in each panel.
}
\label{fig:SF:model}
\end{figure}

The significance of the correct magnetic symmetry in the \ABOshort~compound family lies in its definitive role in the observed complex ME effects.
\CNO has so far been the best studied example. All the components of the ME tensor, connecting the induced electric polarization vector ($P$) to the
magnetic field vector ($H$), have been measured~\cite{Khanh2016}. The experimental observations for
$H\geq$~1~T were explained in the framework of the spin-flop model, in which the magnetic moments are largely perpendicular
to the applied magnetic field.
For $H$ in the [1 1 0] and [1 -1 0] in-plane directions, this model predicts $P$~$\parallel$~[1 1 0].
For $H$~$\parallel$~[0 0 1], the components of $P$ parallel to [1 -1 0] and [0 0 1] are allowed~\cite{Khanh2016}.
That is, the predicted polarization is perpendicular to [1 1 0] in this case.
Detailed description of the symmetry operators is given in Table III.
Experimentally, the largest electric polarization component induced in both $H$~$\parallel$~[1 1 0]
and $H$~$\parallel$~[1 -1 0] is indeed along [1 1 0], in agreement with the model. However, a very significant but unpredicted component of
$P$ is also observed along [1 -1 0]. In Ref.\ \onlinecite{Khanh2016}, this inconsistency was explained by the presence of a minority magnetic
domain with a different direction of the unique monoclinic $\bm{b_m}$ axis. The situation is less satisfactory for $H$~$\parallel$~[0 0 1]. Experimentally,
very similar $P$ components along the allowed [1 -1 0] and the forbidden [1 1 0] directions are observed. This is more difficult to explain by the domain
effects. An alternative scenario was proposed in a more recent single-crystal neutron diffraction study in an applied magnetic
field~\cite{Ding2020:CNO}. According to this work, no spin flop takes place for $H<$ 10 T, and the zero-field magnetic space group is retained.
The induced
electric polarization was proposed to stem from the field-induced redistribution of the magnetic domain populations combined with the tilt
of the magnetic moments, at least in the moderate
fields (below 4 T). Clearly, even the major features of the ME effect in \CNO are far from being understood.

The magnetic structure of \CTO reported in this paper appears to explain its magnetoelectric properties significantly better.
Below, we use the magnetic symmetries of \CTO
to establish the origin of the structure of its ME tensor (allowed and forbidden components),
similar to what has been done for \CNO in Ref.\ \onlinecite{Khanh2016}. The list of the relevant symmetry operators is
given in Table III. Firstly, we note that a spin-flop
transition has been observed in \CTO at $H\sim$ 0.3 T using magnetic susceptibility measurements~\cite{Lee2019}.
Thus, the magnetic moments flop in the [1 -1 0] direction for $H$~$\parallel$~[1 1 0] as shown in Fig.~\ref{fig:SF:model}(a), and largely keep
their directions for $H$~$\parallel$~[1 -1 0], see Fig.~\ref{fig:SF:model}(b). As usual, the moments should slightly tilt in the magnetic field direction.
In the both cases, the $c$ and $c’$ glide planes forbidding the polarization in the [1 1 0] direction are broken,
while the two-fold symmetry axis along [1 1 0] precludes any $P$ normal to this direction.
Therefore, the induced electric polarization along the [1 1 0] direction is expected for both the
$H$~$\parallel$~[1 1 0] and the $H$~$\parallel$~[1 -1 0]
in-plane magnetic fields, see Figs.~\ref{fig:SF:model}(a, b). Experimentally, the polarization of \CTO is indeed along
[1 1 0] for $H$~$\parallel$~[1 -1 0]. For $H$~$\parallel$~[1 1 0], the major component of $P$ is also along [1 1 0], while a significantly smaller [1 -1 0]
component is also present, as in \CNO. The spin-flop model therefore provides a more satisfactory match to the data in \CTO, at least for one of the
in-plane directions.

The magnetic structure of the spin-flop state does not depend on the zero-field easy axis direction for the in-plane magnetic fields, such as
$H$~$\parallel$~[1 1 0] and $H$~$\parallel$~[1 -1 0]. For this reason, the predictions of the spin-flop model are the same for the $C2'/c$ and $C2/c'$
zero-field states and, therefore, for \CTO and \CNO. In contrast, magnetic field along the [0 0 1] direction preserves the in-plane magnetic
structure, and therefore produces different outcomes for $C2'/c$ and $C2/c'$. Fig.~\ref{fig:SF:model}(c) shows the $C2'/c$ magnetic state
(established for \CTO in our paper) for $H$~$\parallel$~[0 0 1]. The $c$ and $c'$ glide planes are broken, and the 2$'$ axis is preserved, allowing
$P$~$\parallel$~[1 1 0] only. As described above, the $C2/c'$ structure reported for \CNO is only compatibe with $P$~$\perp$~[1 1 0].
Clearly, the predictions for the two space groups are mutually exclusive. Experimentally, the
induced polarization in \CTO is essentially along the [1 1 0] direction, with only a small additional [1 -1 0] component. Given the always existing
possibility of the present minority magnetic domain, this is a very satisfactory match to the prediction based on the $C2'/c$ magnetic symmetry.
Based on these observations, one can argue that the results of the ME measurements of Ref.\ \onlinecite{Lee2019} provide an additional independent
confirmation of the $C2'/c$ group in \CTO.

The spin-flop model based on the zero-field $C2'/c$ magnetic order explains the major nonzero terms of the ME tensor of \CTO.
The actual magnetic structure in an applied field could be slightly different because of the (unknown) local magnetic anisotropies. This may
allow additional ME terms, but should not modify the big picture. The larger effect is, probably, the presence of
minority magnetic domains in the samples, as argued in the \CNO literature~\cite{Khanh2016,Ding2020:CNO}. The unexplained polarization along the
[1 -1 0] in the in-plane magnetic field, for example, could be mimicked by a minority [0 1 0]-type domain with the predicted polarization,
coexisting with the majority [1 1 0]-type domain. Clearly, single-domain
samples are crucial for the correct characterization of the ME effects in \CTO and \CNO. One of the important properties of \CTO
is therefore its tendency to form the nearly single-domain state in the flux-grown samples. Without this property, it would be practically
impossible to establish the correct magnetic structure. It would be highly desirable to check the
direction-dependent magnetic and magnetoelectric properties in single-domain samples of \CNO, and determine whether the existing description of this
compound, including its magnetic symmetry, require any modification.

To explain the complex ME properties of \CTO at the lowest temperatures, including the nonlinearity and polarization sign reversal, further
experimental and theoretical work is needed. Characterization of the magnetic order below the anomaly at T~=~6.65~K is an important task for the
future work. Improvements in the consistent mono-domain sample preparation would be crucial for this task. Interestingly, a similar low-temperature
anomaly was found in polycrystalline \MTO~\cite{Panja2019}, suggesting that the Ta ions might play a role in this transition. The peculiar magnetic
and structural symmetries of the \ABOshort~compounds are also expected to give rise to further unusual phenomena, such as quadrupolar excitations
and directional dichroism~\cite{Khanh2016, Matsumoto2019,Cheong2020}. Compounds with the same magnetic point group, such as MnPS$_3$, may
exhibit similar unusual magnetoelectric properties \cite{MnPS3}.
Studies of these phenomena would be of high interest, in our opinion.
First-principles theoretical studies of the \ABOshort~compounds are also highly desired. It was speculated, for instance, that their nonlinear
magnetoelectricity could result from the interplay of the sub-polarizations related to the two inequivalent Co sites~\cite{Solovyev2016, Lee2019}.
The \ABOshort~compound family clearly holds a significant promise for the future work.

\section{Conclusions}
\label{sec:summary}
In conclusion, we report the magnetic order in the magnetoelectric compound \CTO. It consists of collinear and canted antiferromagnetic subsystems,
alternating along the $\bm{c}$ axis. The magnetic space group, $C2'/c$, is different from the one reported previously for this compound, as well as
for the isostructural \CNO. This conclusion was made possible by neutron diffraction, and direction-dependent magnetic susceptibility studies of nearly
mono-domain single crystals. The revised magnetic structure successfully explains the major features of the magnetoelectric effect in \CTO.

\section{Acknowledgments}
We thank Xianghan Xu for the assistance with the traveling solvent floating zone sample growth, and Prof. G. Sch\"utz for supporting the susceptibility
measurements using the MPMS magnetometry. SC thanks Radu Coldea for sharing the software tools for visualizing single-crystal time-of-flight neutron
diffraction data. This work was supported by the NSF under Grant No DMR-1609935. SC was also supported by the international postdoctoral scholarship
at Max Planck Institute for Solid State Research in Germany in the initial stage of this project. Sample growth efforts (JWK and SWC)
were supported by
the DOE under Grant No. DOE: DE-FG02-07ER46382. SP acknowledges financial support from the China Scholarship Council (File No. 201808330123).
The work at Yonsei University was supported by the
National Research Foundation of Korea (NRF) Grants NRF-2017R1A5A1014862 (SRC program: vdWMRC center), NRF-2018R1C1B6006859, and NRF-2019R1A2C2002601.

\appendix

\section{Magnetic susceptibility}
\label{app:sec:az}
\begin{figure}
\includegraphics[width=\linewidth]{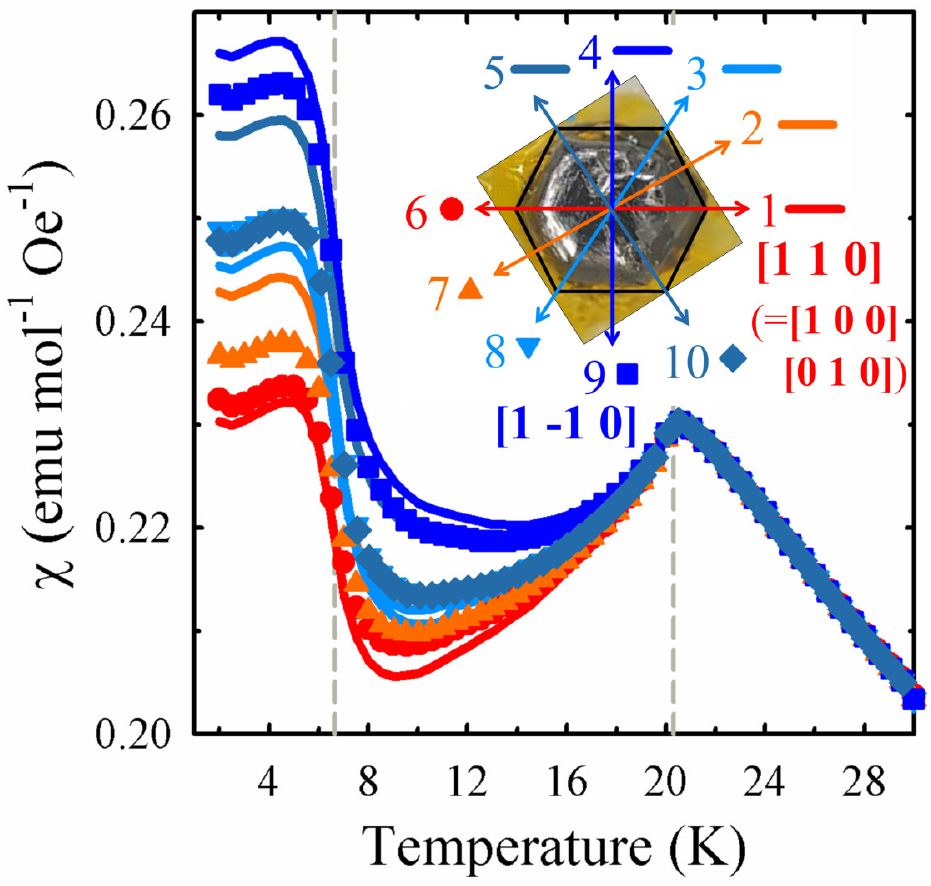}
\caption {(color online) ZFC magnetic susceptibilities for various directions of the $\bm{ab}$-plane magnetic fields, as specified in the inset.
The trigonal notation is used. The data for the opposite magnetic field directions, such as [1 1 0] and [-1 -1 0] are shown with the line and
the symbol of the same color. Vertical dashed lines mark the transition to the long-range magnetic order at 20.3~K, and to the unknown magnetic state
at 6.65~K.
}
\label{fig:SM:rot:Chi:all}
\end{figure}

This Appendix provides details of the direction-dependent ZFC susceptibility measurements related to the reproducibility of the measurements
and the empiric error bars. The determination of the magnetic easy and hard axes in the $\bm{ab}$ plane is of the key importance for this paper.
Therefore, the measurements for each specific direction of the in-plane magnetic field were repeated with the field direction reversed by remounting
the sample. For example, after the
measurement for the [1 1 0] direction of the magnetic field was made, the sample was remounted at room temperature such that the
field was in the opposite [-1 -1 0] direction, and the measurement was repeated. The results of these measurements are expected to be the same, and
therefore any observed discrepancy provides an estimate of the systematic errors.
Fig. \ref{fig:SM:rot:Chi} in the main text shows the averaged results of such measurements. Fig. \ref{fig:SM:rot:Chi:all} shows
all the susceptibility measurements with no averaging. The data for the opposite field directions, such as [1 1 0] and [-1 -1 0], are shown
with the line and the symbol of the same color. The color-coding scheme is chosen such that the directions with the lower susceptibility
are marked with the red-color spectrum, while the higher susceptibility values are depicted using blue colors.
The data of Fig. \ref{fig:SM:rot:Chi:all}
clearly show that while the systematic measurement errors do exist, the conclusion about the hard and the easy magnetic axis is unambiguous.
Importantly, these data prove that the same nearly-monodomain state is formed in the sample on repetitive cooling into the magnetically ordered
state after each sample remounting at room temperature.

\begin{table*}
\caption{Symmetry relations in the two magnetic space group candidates for \CTO. To help visualize the magnetic structure, all the magnetic
moments are expressed using the trigonal axes of the parent structure. The Symmetry column shows how the magnetic moments are generated from the
general atomic position, ($x$, $y$, $z$), using the Seitz notation. The atomic numbering is illustrated in Fig.~\ref{fig:SM:sym}.
The Co ions are in the special
positions with $x$ and $y$ taking the values of 1/3 and 2/3 [see the text for the details]. The chosen magnetic domain has the [0 1 0] easy axis.
This corresponds to the special monoclinic $\bm{b_{m}}$ axis associated with the two-fold and mirror symmetries.
}
\label{tab_sym}
\begin{center}
\begin{tabular}
[c]{c|c|ccc}
\hline\hline
Models          & No. & Coordinates                                        & Moments                    & Symmetry                                  \\
\hline
                      & \#1 & ($x$, $y$, $z$)                                    &(M$_{a}$, M$_{b}$, M$_{c}$)                &\{1$|$0\}                                      \\
$C2/c'$         & \#2 & (1,0,0)+(-$x$, -$x$+$y$, -$z$+1/2)   &(-M$_{a}$, -M$_{a}$+M$_{b}$, -M$_{c}$)     &\{2$_{010}$$|$0 0 1/2\}              \\
(No.~15.88)  & \#3 & (0,1,0)+($x$, $x$-$y$, $z$+1/2)        &(M$_{a}$, M$_{a}$-M$_{b}$, M$_{c}$)          &\{m$^{'}$$_{010}$$|$0 0 1/2\}   \\
                      & \#4 & (1,1,1)+(-$x$, -$y$, -$z$)                   &(-M$_{a}$, -M$_{b}$, -M$_{c}$)            &\{-1$^{'}$$|$0\}                             \\
\hline
                      & \#1 & ($x$, $y$, $z$)                                    &(M$_{a}$, M$_{b}$, M$_{c}$)                &\{1$|$0\}                                        \\
$C2'/c$         & \#2 & (1,0,0)+(-$x$, -$x$+$y$, -$z$+1/2)   &(M$_{a}$, M$_{a}$-M$_{b}$, M$_{c}$)          &\{2$^{'}$$_{010}$$|$0 0 1/2\}      \\
(No.~15.87)  & \#3 & (0,1,0)+($x$, $x$-$y$, $z$+1/2)        &(-M$_{a}$, -M$_{a}$+M$_{b}$, -M$_{c}$)     &\{m$_{010}$$|$0 0 1/2\}              \\
                      & \#4 & (1,1,1)+(-$x$,-$y$, -$z$)                    &(-M$_{a}$, -M$_{b}$, -M$_{c}$)            &\{-1$^{'}$$|$0\}                               \\
\hline\hline
\end{tabular}
\end{center}
\end{table*}

\section{Single-crystal neutron diffraction}
\label{app:ND}
This section provides details of the neutron diffraction data analysis. The structural and magnetic refinements were done using JANA
software~\cite{JANA2016} (partially cross-checked using Fullprof software package~\cite{fullprof}, consistently favoring $C2'/c$). Extinction
corrections were made using the isotropic Becker \& Coppens model implemented in the JANA software~\cite{JANA2016, Becker19}. The absorption
correction was done analytically using the multifaceted crystal model~\cite{Clark1993}. The symmetry analysis to determine the candidate magnetic
space groups using the observed magnetic wave vector $q$~=~(0, 0, 0) was done using Bilbao Crystallographic Server~\cite{Aroyo2011}. The observed
reflections with the intensities $I > 3.0 \times \sigma(I)$ were used in the refinements.

The nuclear structure at 25~K (the paramagnetic state) was refined first, using the entire set of the collected Bragg peaks in the
full range of the scattering vectors $Q$.
This fixed crystal structure was utilized in the magnetic refinements. To attain the reliable magnetic structure determination,
the refinements in the ordered state at 15~K were done using only the lower wave vector peaks with $Q\leq$~6.28~\AA$^{-1}$. This is a standard scheme in
magnetic structure refinement because the magnetic form factor \cite{mag:form:factor:Co2p}
of Co$^{2+}$ goes essentially to zero for $Q>7~\AA ^{-1}$. We found that the standard spin-only magnetic form factor worked well
for our data analysis.
For the systematic analysis, we used the common set of reflections in the refinements at 25 K and 15 K. In addition to the atomic positions
and thermal factors,
all the other parameters related to the nuclear structure refinement, such as the extinction parameters and the scaling factors,
were fixed at their 25 K values in the magnetic refinements.
This approach assumes that the change in the intensities of the Bragg peaks below \textit{T}$_{\rm N}$ is due to the magnetic order.
It is a common choice for a system with
a weak magnetodielectric coupling, as observed in \CNO~\cite{Khanh2019} below \textit{T}$_{\rm N}$, and is justified
by the consistently good quality of the fits.

Overall, 397 (1006) reflections were used in the magnetic refinements for the flux (TSFZ) sample. These are the reflections shown in
Fig.~\ref{fig:SM:ND:flux} in the main text.
To be consistent with the magnetic susceptibility data, the magnetic moments in the $C2'/c$ refinements may exhibit only a small tilt away
from the crystallographic $\bm{b}$ axis. Unrestricted refinements generally did not meet this condition, and therefore models with correlated
tilting of the spins away from the $\bm{b}$ axis were considered.
Tilting the adjacent Co1 and Co2 moments in the opposite directions
produced better results than tilting them together in the same direction. The best results were achieved when the M$_a$ components
of Co1 and Co2 were constrained to take the opposite values, while the M$_b$ components were kept equal.
The values of the Co$^{2+}$ magnetic moments shown in Table~\ref{tab:mag:ref} are systematically
smaller for the flux crystal in comparison to those of the TSFZ sample. This discrepancy is slightly reduced if multi-domain refinements are done
for the flux crystal. Since the quality of the fit was not improved significantly in such trials, and because our magnetic susceptibility
measurements consistently identify the flux crystal being close to the dominant single-domain state, these refinements were not pursued further.
We note that the observed variation of the magnetic moment values lies well within the limits of the discrepancy of the reported moments for
\CNO~\cite{Khanh2016,Deng2018,Ding2020:CNO}, and is therefore attributable to the general accuracy of the method.

\begin{figure} [t!]
\includegraphics[width=\linewidth]{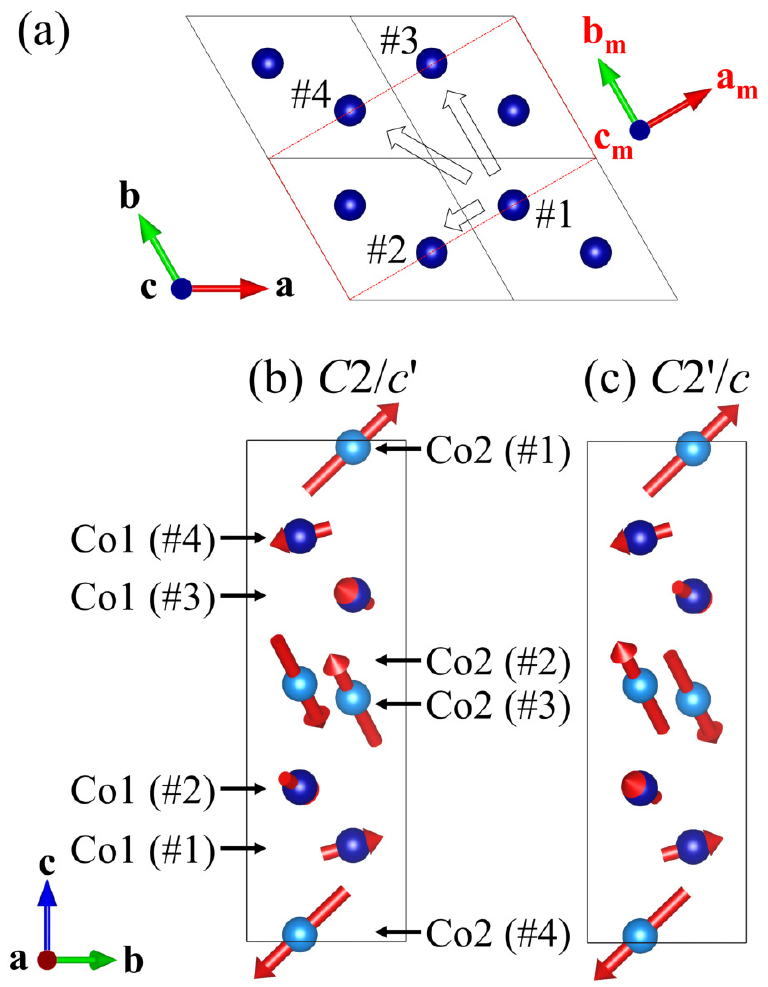}
\caption {(color online) (a) Symmetry relations between the parent trigonal crystallographic axes, and the magnetic monoclinic axes.
The latter are marked with subscript $\bf{m}$. The trigonal and monoclinic unit cells are shown with solid black lines and dotted red lines,
respectively. The numbered atoms are connected by the symmetry transformations given in Table~\ref{tab_sym}, as marked with empty white arrows.
(b, c) Example magnetic structures generated by $C2/c'$ and $C2'/c$. We arbitrarily chose ${\bm M}$(Co1)~=~(0.2, 0.4, 0.1)~$\mu_{B}$, ${\bm M}$(Co2)~=~(0.3, 0.6, 0.5)~$\mu_{B}$
to generate both magnetic structures.
Note that these general examples
do not correspond to the actual magnetic structure of \CTO determined in this work.
}
\label{fig:SM:sym}
\end{figure}

\begin{figure*} [t!]
\includegraphics[width=\linewidth]{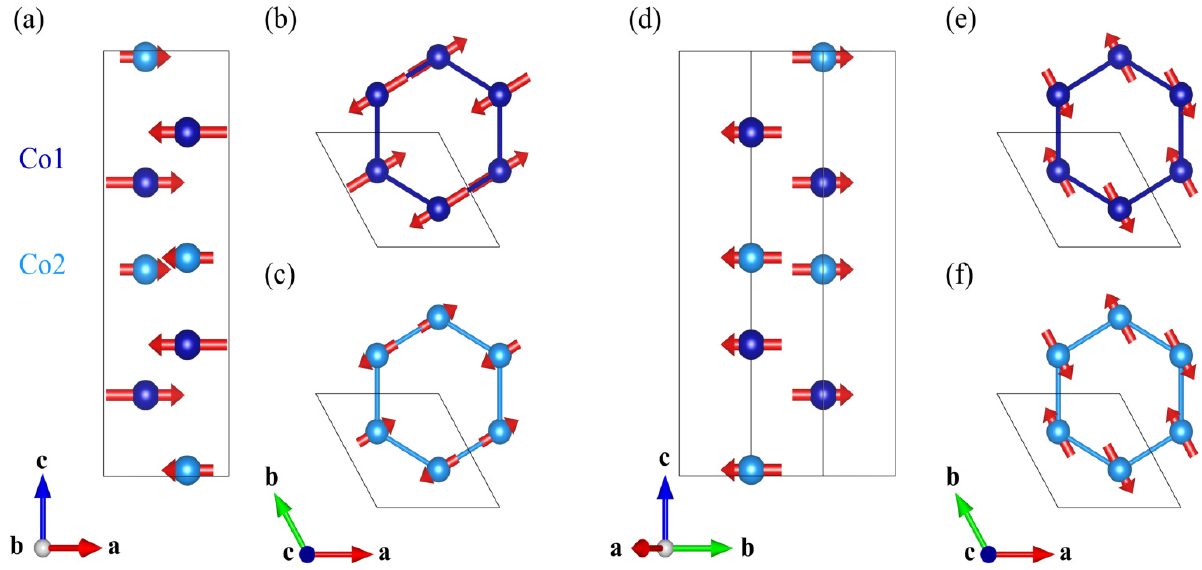}
\caption {(color online)
In-plane collinear antiferromagnetic order generated by the $C2/c'$ group (a-c), and by the $C2'/c$ group (d-f), in the
trigonal unit cell. The magnetic moment values are from the TSFZ data analysis given in Table I.
}
\label{fig:SM:collinear}
\end{figure*}

\section{Magnetic symmetry relations}
\label{app:MSG}
In this work, the magnetic refinements were carried out using two candidate monoclinic magnetic space groups, $C2/c'$ and $C2'/c$. As the magnetic
order in the \ABOshort~compounds can be visualized better using the unit cell and the crystallographic axes of the parent trigonal space group \sg, we
consistently adopted the trigonal setting in this paper. This Appendix provides the relevant symmetry relations for those magnetic space groups.
The parent trigonal \CTO structure has two crystallographically different Co sites, denoted as Co1 and Co2. These distinct sites form the buckled
and the flat honeycomb Co layer, respectively. They occupy the $4d$ Wyckoff position, generating four equivalent sites: (1/3, 2/3, $z$), (2/3, 1/3, -$z$+1/2),
(2/3, 1/3, -$z$), and (1/3, 2/3, $z$+1/2). The symmetry operators connecting the magnetic moments are given using the Seitz notation in Table~\ref{tab_sym}. This
notation gives the rotational transformation on the left, and the following translational transformation on the right. The atoms connected by these
transformations are marked in Fig.~\ref{fig:SM:sym}(a), the white arrows depict the transformations. The relations between the crystallographic axes of the trigonal and
monoclinic unit cells are shown in Fig.~\ref{fig:SM:sym}(a). The unit cell axes are transformed as ${\bm a_{m}}=2\bm{a}+\bm{b}$, ${\bm b_{m} }=\bm{b}$,
${\bm c_{m}}={\bm c}$, where the subscript $m$ refers to the monoclinic axes. Table~\ref{tab_sym} provides the magnetic moments in the trigonal
notation, and therefore it can be used directly to generate the magnetic structures in the parent unit cell. As an example, two magnetic
structures generated using these rules for the $C2/c'$ and the $C2'/c$ groups are shown in Figs.~\ref{fig:SM:sym}(b, c).
The atomic numbering is consistent with Table II. For the buckled layer, $z$~$\sim$~0.192, while for the flat layer $z$~$\sim$~0.986 (room temperature
values~\cite{Lee2019}).
These figures also illustrate that the net magnetic moment of the unit cell is zero even for a general noncollinear structure.
Note that these magnetic structures are of the general type, unrelated to the actual structure of \CTO.

Regarding \CTO, the important realizations of these symmetry rules are the collinear structures with zero $\bm{c}$-axis moments.
For the collinear structures consistent with the neutron difraction data,
Table~\ref{tab_sym} gives M$_{a}$ = 2 M$_{b}$ for $C2/c'$, and M$_{a}$ = 0 for $C2'/c$. The corresponding magnetic structures are illustrated in
Fig.~\ref{fig:SM:collinear}. Note that the magnetic moments in one of these structures (and therefore the corresponding magnetic easy axis) are
perpendicular to the moments in the other. The origin of the three in-plane magnetic domains for $C2'/c$ can be best understood using the structure shown in
Figs.~\ref{fig:SM:collinear} (e, f) and the crystallographic axes in Fig.~\ref{fig:SM:sym}(a). In both figures, the moments are along the coinciding
trigonal $\bm{b}$ and monoclinic $\bm{b_m}$ axes. The other two magnetic domains are generated when the $\bm{b_m}$ axis points along
the equivalent [1 0 0] or [-1 -1 0] trigonal direction. The latter domain, with the easy axis along the [1 1 0], and the
hard axis along the [1 -1 0] is usually used in the literature describing the magnetoelectric effects in this family of compounds~\cite{Lee2019}.

\begin{table}
\caption{Symmetry operators of \CNO and \CTO.
For $H$~$\parallel$~[1 1 0] and $H$~$\parallel$~[1 -1 0], they are applicable to the spin-flop model. For $H$~$\parallel$~[0 0 1],
the operators are listed for the published $C2/c'$ group for \CNO, and for the $C2'/c$ group determined in this paper for \CTO.
$2$ is the two-fold symmetry axis along the [1 1 0], $c$ is the [1 1 0] glide plane.
The prime symbol ($'$) denotes the time reversal operation.
}
\label{tab:sym:sf}
\begin{center}
\begin{tabular}
[c]{|c|c|c|c|c|}
\hline
                                                    & \multirow{2}{*}{Symmetry}    &\multirow{2}{*}{H\,$\parallel$\,[1 1 0]}     &\multirow{2}{*}{H\,$\parallel$\,[1 -1 0]}  &\multirow{2}{*}{H\,$\parallel$\,[0 0 1]}  \\
                                                    &                                                &                                                 &                                                &                                               \\
\hline
{\CNO}                                         &     broken           &$2'$, $c$, $c'$      &$2$, $c$, $c'$        &$2$, $2'$, $c$    \\ [0.15cm]
(Ref.\ \onlinecite{Khanh2016})  &     allowed          &$2$                         &$2'$                        &$c'$                      \\
\hline
{\CTO}                                         &     broken           &$2'$, $c$, $c'$      &$2$, $c$, $c'$        &$2$, $c$, $c'$              \\ [0.15cm]
(this work)                                   &     allowed          &$2$                         &$2'$                        &$2'$                       \\
\hline
\end{tabular}
\end{center}
\end{table}

For completeness, we provide the symmetrically-allowed ME tensors of $C2/c'$ and $C2'/c$ in zero field~\cite{Aroyo2011} and in the spin-flop
phase [see Table~\ref{tab:ME}]. The ME tensor is defined as $P_{i}$ = $\alpha_{\textit{ij}}$ $H_{j}$ where
$P_{i}$ ($H_{j}$) is the electric polarization (the magnetic field) along the principal direction $i$ ($j$). The principal axes are
[1 1 0], [1 -1 0], and [0 0 1] in the trigonal lattice, according to the conventional crystallographic definition of the ME tensor for the
non-orthogonal crystallographic lattice.

\begin{table}
\caption{The ME tensors ({$\alpha_{\textit{ij}}$}) of the $C2/c'$ and $C2'/c$ monoclinic magnetic space groups, for zero magnetic field, and
in the spin-flop state.}
\label{tab:ME}
\begin{center}
\begin{tabular}{|*3{>{\centering\arraybackslash}p{.15\textwidth}|}}
\hline
& \multirow{2}{*}{Ambient state} & \multirow{2}{*}{Spin-flop state} \\
& & \\
\hline
\multirow{6}{*}{$C2/c'$ (No.~15.88)} & \[ \left[ \begin{array}{ccc} \alpha_{\textit{11}} & 0 & \alpha_{\textit{13}} \\ 0 & \alpha_{\textit{22}} & 0 \\ \alpha_{\textit{31}} & 0 & \alpha_{\textit{33}} \end{array}\right]\]
                                                             & \[ \left[ \begin{array}{ccc} \alpha_{\textit{11}} & \alpha_{\textit{12}} & 0 \\ 0 & 0 & \alpha_{\textit{23}}  \\ 0 & 0 & \alpha_{\textit{33}} \end{array}\right]\]  \\
\hline
\multirow{6}{*}{$C2'/c$ (No.~15.87)} & \[ \left[ \begin{array}{ccc} 0 & \alpha_{\textit{12}} & 0 \\ \alpha_{\textit{21}} & 0 & \alpha_{\textit{23}} \\ 0 & \alpha_{\textit{32}} & 0 \end{array}\right]\]
                                                             & \[ \left[ \begin{array}{ccc} \alpha_{\textit{11}} & \alpha_{\textit{12}} & \alpha_{\textit{13}} \\ 0 & 0 & 0 \\ 0 & 0 & 0 \end{array}\right]\]  \\
\hline
\end{tabular}
\end{center}
\end{table}

\section{Magnetic structure factor of (0, 0, L) peaks}
\label{app:MSF:001}
Herein, the analytical expression for the magnetic structure factor of the (0, 0, L) peaks is given for the $C2'/c$ and $C2/c'$ magnetic space
groups. The magnetic structure factor is defined as
\begin{equation}
\bm F_{hkl}=\sum_{\rm{i=1}}^{N} {\bm M}_i e^{i {\bm k} \cdot {\bm R}_{i}}, \label{mag:str:fac}
\end{equation}
where ${\bm M}_i$ is the magnetic moment vector at the $i$-th ionic site, ${\bm R}_{i}$ is the fractional coordinate of the $i$-th ion in the unit
cell. $N$ runs from 1 to 4 as there are four equivalent atomic sites in these magnetic space groups. The magnetic structure factors of (0, 0, L)
can be calculated using the atomic coordinates and the magnetic moments in Table~\ref{tab_sym}. One obtains
\begin{equation}
\bm F_{(0,0,{\rm L})} = -2 i \sin(2 \pi z {\rm L}) ({\rm M}_{a} - 2 {\rm M}_{b}) \hat{b}
\label{mag:str:F001:15p88}
\end{equation} \\
for $C2/c'$ (No.~15.88) and
\begin{equation}
\bm F_{(0,0,{\rm L})} = 2 i \sin(2 \pi z {\rm L}) (2 {\rm M}_{a}\hat{a} + {\rm M}_{a}\hat{b} + 2 {\rm M}_{c}\hat{c})
\label{mag:str:F001:15p87}
\end{equation}
for $C2'/c$ (No.~15.87). Here $\bm{M}$ = (M$_{a}$, M$_{b}$, M$_{c}$) is the magnetic moment of the \# 1 site in the trigonal notation,
$z$ is the fractional coordinate of the \# 1 site for either Co1 or Co2, and {\textit{\^{a}}}, {\textit{\^{b}}}, {\textit{\^{c}}} are the trigonal crystallographic unit
vectors. Equations~(\ref{mag:str:F001:15p88}) and~(\ref{mag:str:F001:15p87}) are identical for both Co1 and Co2 sites. The net magnetic
structure factor is the sum of the contributions from all the Co sites in the unit cell, and the magnetic Bragg peak intensity~\cite{Squires1996} is
proportional to $\mid$$\bm F_{(0,0,{\rm L})}$$\mid$$^{2}$.

The (0, 0, odd) nuclear Bragg peaks are forbidden in the parent trigonal structure, as well as in the monoclinic space groups considered here.
For the collinear magnetic order,
M$_{a}$ = 2 M$_{b}$ in $C2/c'$, and $\bm{M}$ = M$_{b}$ \^{b} in $C2'/c$, as discussed above. In these cases,
equations~(\ref{mag:str:F001:15p88}) and~(\ref{mag:str:F001:15p87}) give zero intensities for any odd L. When the above conditions are broken
and the magnetic moments become noncollinear, non-zero (0, 0, odd) structure factors are obtained. The moments from both the buckled and flat
honeycomb layers contribute to the peak intensity. Importantly, only the magnetic moment components perpendicular
to the scattering wave vector $Q$ are used in the calculation of the structure factor in neutron diffraction~\cite{Squires1996}.
For the (0, 0, L) reflections, these are
the $\bf{ab}$-plane components. Therefore, our experimental observation of the (0, 0, odd) magnetic peaks directly proves that the magnetic
structure is noncollinear in the $\bf{ab}$ plane.

\end{document}